\documentclass[sigconf, screen, anonymous=False]{acmart}
\PassOptionsToPackage{dvipsnames}{xcolor}
\PassOptionsToPackage{colorlinks=true,linkcolor=RubineRed,breaklinks=True,citecolor=RubineRed}{xcolor}

\usepackage{enumitem}% http://ctan.org/pkg/enumitem
\setlist[itemize]{noitemsep, topsep=0pt}

\usepackage{tabularray}
\usepackage[most]{tcolorbox}
\usepackage{adjustbox}
\usepackage[linesnumbered,algo2e,ruled]{algorithm2e}
\usepackage{multirow}
\usepackage{booktabs,subcaption,dcolumn}
\usepackage{tikz}
\usepackage{enumitem}
\usepackage{tabularx}
\usepackage{MnSymbol}
\usepackage{makecell} % in preamble
\usepackage{pifont}% http://ctan.org/pkg/pifont
\usepackage{svg}
\usepackage{soul}
\usepackage{amsmath}
\usepackage{dirtytalk}
\usepackage{subcaption}
\usepackage{graphicx}
\usepackage{setspace}
\usepackage[font=small]{caption}
\usepackage{booktabs} %toprule / bottomrule 
\usepackage[titletoc,toc,title]{appendix}
\usepackage{svg}
\usepackage{amsfonts}
\usepackage{xurl} % split urls too long

\usepackage{diagbox}%y 18 mar
\usepackage{indentfirst}
\usepackage{listings}
\usepackage{xcolor}
\lstdefinestyle{slrquery}{
  basicstyle=\ttfamily\footnotesize,
  frame=single,
  framerule=0.4pt,
  breaklines=true,
  breakatwhitespace=true,
  columns=fullflexible,
  keepspaces=true,
  aboveskip=0.5\baselineskip,
  belowskip=0.5\baselineskip,
  numbers=none,
  showstringspaces=false
}
\usepackage{CJKutf8}
\usepackage{amsmath}
\usepackage{amsfonts} %, amssymb}

\usepackage{colortbl}

\copyrightyear{2026}
\acmYear{2026}
\setcopyright{cc}
\setcctype{by}
\acmConference[CODASPY '26]{Proceedings of the Sixteenth ACM Conference on Data and Application Security and Privacy}{June 23--25, 2026}{Frankfurt am Main, Germany}
\acmBooktitle{Proceedings of the Sixteenth ACM Conference on Data and Application Security and Privacy (CODASPY '26), June 23--25, 2026, Frankfurt am Main, Germany}
\acmDOI{10.1145/3800506.3803505}
\acmISBN{979-8-4007-2562-3/2026/06}

\usepackage[dvipsnames]{xcolor}

\AtBeginDocument{%
  \providecommand\BibTeX{{%
    \normalfont B\kern-0.5em{\scshape i\kern-0.25em b}\kern-0.8em\TeX}}}


\newcolumntype{?}{!{\vrule width 1.5pt}}

\newtcolorbox{cooltextbox}[1][]{%
    colback=black!5,
    colframe=black!5,
    notitle,
    sharp corners,
    borderline west={0pt}{0pt}{red!80!black},
    enhanced,
    breakable,
    left=0pt,
    right=0pt,
    top=0pt,
    bottom=0pt
    }

\newcommand\revision[1]{%
  \bgroup
  \hskip0pt\color{blue!80!black}%
  #1%
  \egroup
}

\begin{document}
% \title{Practical Barriers to Trustworthy AI-Driven Threat Intelligence in Finance: A User-Centric Study}

\title{Security Barriers to Trustworthy AI-Driven Cyber Threat Intelligence in Finance: Evidence from Practitioners}
% \title{Beyond Accuracy: Practical Barriers to Trustworthy AI-Driven Threat Intelligence in Finance}
% \title{Trust, Usability, and Fragility: Barriers to Deploying AI-Driven Threat Intelligence in Finance}

% \titlerunning{Use this for LNCS if the title is too long}

\thispagestyle{empty}
% \titlenote{Produces the permission block, and copyright information}
\author{Emir Karaosman}
\email{emir.karaosman@uni.li}
\affiliation{%
%  \department{Liechtenstein Business School}
  \institution{University of Liechtenstein}
  \city{Vaduz}
  \country{Liechtenstein}
}

\author{Advije Rizvani}
\email{advije.rizvani@uni.li}
\affiliation{%
%  \department{Liechtenstein Business School}
  \institution{University of Liechtenstein}
  \city{Vaduz}
  \country{Liechtenstein}
}

\author{Irdin Pekaric}
\email{irdin.pekaric@uni.li}
\affiliation{%
%  \department{Liechtenstein Business School}
  \institution{University of Liechtenstein}
  \city{Vaduz}
  \country{Liechtenstein}
}

\pagestyle{empty}

\begin{abstract}

Financial institutions face increasing cyber risk while operating under strict regulatory oversight. To manage this risk, they rely heavily on Cyber Threat Intelligence (CTI) to inform detection, response, and strategic security decisions. Artificial intelligence (AI) is widely suggested as a means to strengthen CTI. However, evidence of trustworthy production use in finance remains limited. Adoption depends not only on predictive performance, but also on governance, integration into security workflows and analyst trust. Thus, we examine how AI is used for CTI in practice within financial institutions and what barriers prevent trustworthy deployment.

We report a mixed-methods, user-centric study combining a CTI-finance-focused systematic literature review, semi-structured interviews, and an exploratory survey. Our review screened 330 publications (2019--2025) and retained 12 finance-relevant studies for analysis; we further conducted six interviews and collected 14 survey responses from banks and consultancies. Across research and practice, we identify four recurrent socio-technical failure modes that hinder trustworthy AI-driven CTI: (i) \emph{shadow use} of public AI tools outside institutional controls, (ii) \emph{license-first enablement} without operational integration, (iii) \emph{attacker-perception gaps} that limit adversarial threat modeling, and (iv) \emph{missing security for the AI models themselves}, including limited monitoring, robustness evaluation and audit-ready evidence. Survey results provide additional insights: 71.4\% of respondents expect AI to become central within five years, 57.1\% report infrequent current use due to interpretability and assurance concerns and 28.6\% report direct encounters with adversarial risks. Based on these findings, we derive three security-oriented operational safeguards for AI-enabled CTI deployments and release supporting materials (search strings, coded evidence matrix, and interview and survey instruments) to improve transparency and facilitate replication and adaptation in similar organizational settings.

\end{abstract}

%    <concept>
%        <concept_id>10010147.10010178.10010179</concept_id>
%        <concept_desc>Computing methodologies~Natural language processing</concept_desc>
%        <concept_significance>500</concept_significance>
%        </concept>
%    <concept>
%        <concept_id>10002978.10003029</concept_id>
%        <concept_desc>Security and privacy~Human and societal aspects of security and privacy</concept_desc>
%        <concept_significance>500</concept_significance>
%        </concept>
%  </ccs2012>

% \ccsdesc[500]{Computing methodologies~Natural language processing}
% \ccsdesc[500]{Security and privacy~Human and societal aspects of security and privacy}

% \keywords{Trustworthy AI, Cyber Threat Intelligence, Adoption Barriers, Financial Sector, AI Explainability
% }

\settopmatter{printfolios=false}

\maketitle

\section{Introduction}
\label{sec:introduction}
\noindent

\noindent Financial institutions operate in a fast-moving threat landscape while meeting strict risk and compliance expectations. This makes finance particularly critical as it underpins global economies. Breaches such as the 2017 Equifax attack \cite{Fruhlinger2020Equifax}, which affected 147 million users and cost over \$1.4 billion, show how technical weaknesses in finance ripple across entire economies. As data flows, algorithms, and automation grow more complex, institutions face an ongoing tension: \textbf{how to innovate securely while maintaining compliance and trust?} Within this environment, Artificial Intelligence (AI) and Machine Learning (ML) have emerged as potential accelerators of Cyber Threat Intelligence (CTI), offering faster detection, broader coverage, and data-driven insight \cite{cyberthreatsclassif, kaloudi2020ai, ceccarelli2014continuous, viegas2017toward}.

AI is transforming CTI by automating pattern recognition, highlighting anomalies, and correlating weak signals across heterogeneous sources [15]. Its goal is to anticipate emerging threats earlier and reduce the burden on analysts [33]. Security teams rely on both public and private information sources, yet integrating these into actionable knowledge remains a persistent challenge [Felderer and Pekaric, 2017].

AI is transforming CTI by automating pattern recognition, highlighting anomalies and correlating weak signals across heterogeneous sources \cite{cybersecurityinfinancialservices}. Its goal is to anticipate emerging threats earlier by integrating public and private data sources into actionable knowledge to reduce the burden on analysts \cite{pfister2025department, felderer2017research}. Yet, despite heavy investment and enthusiasm, most financial institutions still rely on conventional rule-based systems, complemented by human triage \cite{alam2023looking, groner2023model}. AI adoption remains sporadic, as accuracy alone has proven insufficient to make systems operationally credible. Explainability, auditability, and regulatory defensibility---not merely model performance---determine whether AI-generated intelligence is trusted \cite{IIFMcKinsey2024, pekaric2026llms}. Recent studies show that ML-based security models often fail under distribution shifts or adversarial inputs, raising questions about confidence in AI-driven outputs \cite{herd2024can, rizvani2025ephemeral}. Despite growing interest, real-world adoption remains limited: prior research offers many models and benchmarks for security tasks, but evidence of production use in finance is scarce \cite{bigdataplatformforintegrated, designoffinancialnetwork}. Studies often optimize on public or retrospective datasets and do not discuss how outputs fit analyst decision making, how systems satisfy model risk management and audit requirements, and how they withstand adversarial manipulation and misuse \cite{9944106, castro2019aimed, rossolini2023real, araujo2025enhancing}. Finance-specific CTI tools and datasets exist, yet they rarely connect model performance to analyst trust, integrate into operational CTI workflows, or align with regulatory controls \cite{cyberseucirtyindigitaltransformation}.

\textbf{Research gap and RQs.}
This gap between AI-driven CTI research and its trustworthy and regulated deployment in financial institutions motivates a closer look at how AI is actually used for CTI in finance today, what limits adoption, and which safeguards improve trust and robustness. We begin with a scoping question that motivates our systematic literature review: \textbf{RQ0 (coverage):} \textit{To what extent does existing research address trustworthy and deployable AI-driven CTI in finance, beyond isolated task performance?} Building on this foundation, we study three additional questions: \textbf{RQ1 (adoption):} \textit{How is AI currently adopted for CTI in financial institutions?} \textbf{RQ2 (barriers):} \textit{What practical barriers limit adoption and operational use?} \textbf{RQ3 (trust and robustness):} \textit{What drives (or undermines) trust in AI-driven CTI outputs, and how are robustness and misuse risks handled in practice?}

\textbf{Method.}
To address these questions, we combine a systematic literature review covering \textit{2019 to 2025} (§\ref{ssec:slr}) with practitioner evidence from semi-structured interviews (§\ref{ssec:interviews}) and an exploratory survey (§\ref{ssec:surveys}). The review retains \textit{12} finance-relevant publications that involve AI or critically analyze AI-related risks within CTI. The interviews span multiple organizations in banking and consulting and include senior roles such as Head of Cyber Defense, Information Security Officer, and Cyber and Digital Risk Specialist. The survey provides additional triangulation of adoption signals and barriers in current practice. We code all evidence against the research questions above (§\ref{ssec:problem}).

\textbf{Synthesis of observations.}
Across both the literature and practice, we find that the core challenges in adopting AI for CTI in finance are socio-technical rather than purely technical. Four recurring patterns define the current landscape: (1) \textit{shadow AI in the wild}, where employees use public large-language models despite institutional restrictions; (2) \textit{a license-first trap}, where adoption follows vendor bundling rather than clear capability needs; (3) \textit{attacker-perception gaps}, as defenders underestimate offensive capabilities of adversaries; and (4) \textit{missing security for AI itself}, with minimal drift monitoring, adversarial testing, or audit-ready documentation. Altogether, these findings show that while AI models in finance are becoming more capable, institutional governance, explainability, and accountability have not evolved at the same pace. As a result, AI deployments that appear promising in theory often fail to meet the operational, regulatory, and trust requirements of financial cybersecurity.

\noindent\textbf{CONTRIBUTIONS.} We advance the understanding of AI enabled CTI in finance by:
\begin{itemize}[leftmargin=*,noitemsep,topsep=0pt]
  \item \textbf{A systematic synthesis of prior research.} We show that prior work largely focuses on isolated, task-level performance and overlooks deployment-critical aspects of financial CTI. In this regards, we review 330 publications (2019–2025) and retain 12 finance-relevant studies, which show that while AI improves accuracy in isolated tasks such as fraud and phishing detection, there are only few works that address explainability, human-AI collaboration, or resilience against adversarial manipulation. This represents the first domain-specific map of what ``AI in financial CTI'' currently covers as well as where it remains incomplete.
  \item \textbf{An empirical, mixed-methods and multi-organizational investigation grounded in practice.}
    We combine six semi-structured interviews and fourteen survey responses from banks and consultancies to examine how AI tools are actually deployed and constrained in financial institutions. This is the first empirical combination of these methods applied to AI adoption in financial CTI. Our findings expose barriers related to data integration, procurement dependencies, regulatory defensibility and limited hybrid expertise.
  \item \textbf{Actionable recommendations grounded in our findings.} We translate empirical insights into concrete guidance for practitioners and policymakers. This is supported by practitioner-ready artifacts for replication and further study, including search strings, a coded evidence matrix, and interview and survey instruments.
\end{itemize}

To support transparency and reuse, supplementary materials\footnote{We release the full SLR protocol, including search strings, screening criteria, and a coded evidence matrix, enabling independent re-execution of the review process. Application of the same protocol should yield a comparable evidence base. The released interview and survey instruments further support replication and extension of the empirical design in similar organizational settings.} are provided in our repository: \url{https://github.com/irdin-pekaric/CODASPY2026}.

\section{Related Work and Motivation}
\label{sec:related}
\noindent

\noindent We summarize the use of AI in finance, and briefly discuss the use of AI in CTI (§\ref{ssec:background}). Then, we present our systematic literature review (§\ref{ssec:slr}), and define the research gap (§\ref{ssec:problem}).

\subsection{Background}
\label{ssec:background}

\noindent Financial institutions devote many resources to defend against fraud, market abuse, and cyber threats while operating under strict regulatory and audit requirements \cite{sotgiu2022explainability,casola2020novel}. Yet, static rule-based controls and traditional monitoring often fail to keep pace with adaptive attackers, creating blind spots in CTI workflows \cite{zoppi2021unsupervised, campos2022online, cullen2022adversarial}. Consequently, AI and ML are increasingly explored as enablers to enhance detection, streamline CTI, and support security operations in finance \cite{catillo2024successful, andresini2022roulette, zoppi2021meta, badjie2024denoising}.

\subsubsection*{AI in Finance}
AI and ML have moved from experimental tools to core enablers in financial services. Banks and other institutions increasingly rely on them for fraud detection, transaction monitoring, market-abuse detection, and anti-money laundering \cite{cybersecurityinfinancialservices,creditcardfraud,cyberthreatsclassif}. In controlled experiments, supervised and deep learning approaches frequently outperform rule-based systems on task-specific metrics, particularly in tasks such as card-fraud screening, phishing-activity forecasting, and network-anomaly detection \cite{creditcardfraud,utilizingdeeplearning,designoffinancialnetwork}.
Beyond individual models, researchers have begun to propose domain-specific platforms that fuse multiple sources and provide sector-wide situational awareness, for example through extensions of open standards for structured CTI sharing to financial infrastructures \cite{bigdataplatformforintegrated,designoffinancialnetwork}. Yet, translating technical promise into practice remains difficult. At the same time, deployment decisions in finance are governed by regulatory expectations around explainability, auditability, and model risk management rather than predictive performance alone \cite{cybersecurityinfinancialservices,cyberthreatsclassif,aihypeasacybersecurityrisk}. Security tools must produce outputs that can be justified in internal reviews and external audits, integrated into existing Security Operations Center (SOC) workflows, and defended under regulatory scrutiny. These constraints directly shape how AI systems are evaluated and adopted in practice, and later manifest as barriers to trust and operational deployment (see §\ref{ssec:rq2}, §\ref{ssec:rq3}).

\subsubsection*{AI in Cyber-Threat Intelligence}
CTI involves the collection, analysis, and dissemination of information about adversaries, indicators, and tactics to support detection, response, and strategic security decision making. In this context, AI is increasingly explored as a force multiplier that can reduce analyst workload and help triage large volumes of heterogeneous CTI inputs. Concretely, machine learning models have been applied to mine structured and unstructured sources, forecast threat surges such as phishing campaigns, and enrich SOC alerts with contextual information. However, the adoption of AI in CTI brings its own challenges. To be trusted in operational workflows, systems must not only provide accurate outputs but also explanations that analysts can interpret; integration must be feasible within existing security operations centers; and the models themselves must withstand adversarial manipulation \cite{attackercentricthinking,aihypeasacybersecurityrisk,cybersecurityinfinancialservices}.
In short, while AI extends CTI’s reach, its practical value depends on overcoming barriers of trust, integration, and robustness \cite{attackercentricthinking,cybercrimethreatintelligence,aihypeasacybersecurityrisk}. As a result, the literature offers limited insight into why AI remains peripheral in production CTI workflows.

\subsection{Systematic Literature Review}
\label{ssec:slr}

\noindent As a starting point of our research, we asked ourselves a preliminary research question (RQ0): \textcolor{violet!90!black} {\textit{to what extent has prior research addressed a trustworthy use of AI for CTI in financial institutions, including technical efficacy, human factors, integration, and exposure to adversarial risks?}}

\subsubsection*{Methodology}
Our SLR is based on established practices: we first collect relevant papers, and then analyse them~\cite{rethlefsen2021prisma}. This is done in line with previous systematic reviews in security domains~\cite{pekaric2023systematic}. In the following, we describe the systematic protocol we applied to carry out each of these phases (carried out in Q1 2025).

\subsubsection*{Collection phase}
We conducted a systematic search by following the guidelines by Kitchenham et al.~\cite{kitchenham2015evidence}. The protocol specifies (i) database selection, (ii) search strategy, (iii) inclusion and exclusion criteria, (iv) a staged screening process and (v) a structured data extraction and coding phase. We use keyword-based searches across five scholarly databases: \textit{IEEE Xplore}, \textit{ACM Digital Library}, \textit{SpringerLink}, \textit{Elsevier ScienceDirect}, \textit{Association for Information Systems (AIS)}; as well as proceedings from \textit{NDSS Symposium} and \textit{USENIX Security Symposium}. Queries were centered across three main keywords: finance, CTI, and AI. These were combined with task-specific keywords (e.g., fraud, phishing, SIEM, threat hunting) to search forpapers published in the 2019--2025 window. For reproducibility, the exact search string\footnote{We discuss the limitations of our search queries in §\ref{ssec:limitations}.} is reported in Listings~\ref{lst:long}.

The search yielded 330 records. After deduplication and multi-stage screening (title $\rightarrow$ abstract $\rightarrow$ full text), 12 peer-reviewed publications\footnote{The small number of included papers is due to the intersectional focus on finance-specific, CTI-relevant and AI-enabled studies. This indicates a sparse evidence base and not the reviewer bias.} were retained for analysis (see Appendix~\ref{app:first}, Table~\ref{tab:slr_search_results} for per-database counts). Inclusion and exclusion criteria were applied progressively during screening. Inclusion required:
\begin{itemize}[leftmargin=*,noitemsep,topsep=0pt]
  \item a clear finance context or evaluation within financial institutions;
  \item use or critical analysis of AI/ML at any CTI stage (collection, processing, analysis, dissemination);
  \item or explicit treatment of AI-specific risks (e.g., adversarial manipulation, model trust).
\end{itemize}
We excluded non-peer-reviewed work, non-English texts, crypto-only studies lacking CTI scope, and papers with no focus on cybersecurity. These criteria were enforced during abstract and full-text screening --- papers failing any criterion at a given stage were excluded and not reconsidered at later stages.

\begin{lstlisting}[style=slrquery,caption={Search String},label={lst:long}, float, floatplacement=t]
("finance" OR "bank" OR "financial services")
    AND ("cyber threat intelligence" OR "CTI" OR "threat intel")
    AND (Abstract:("AI") OR Abstract:("machine learning") OR Abstract:("deep learning") OR Abstract:("LLM") OR Abstract:("generative"))
    AND (Abstract:("fraud") OR Abstract:("phishing") OR Abstract:("SIEM") OR Abstract:("threat hunting"))
\end{lstlisting}

\begin{table}[t]
  \centering

  \scriptsize
  \setlength{\tabcolsep}{2pt}
  \renewcommand{\arraystretch}{1.08}
  \newcolumntype{Y}{>{\raggedright\arraybackslash}X}
  \newcolumntype{C}[1]{>{\centering\arraybackslash}p{#1}}

  \begin{tabularx}{\linewidth}{@{}%
     Y
     C{0.9cm}
     Y
     @{\hspace{2pt}}
     C{0.85cm}
     C{0.9cm}
     C{1.15cm}
  @{}}
    \textbf{Authors (short title)} & \textbf{Year} & \textbf{Type of study} &
    \textbf{\shortstack{RQ1\\(adoption)}} &
    \textbf{\shortstack{RQ2\\(barriers)}} &
    \textbf{\shortstack{RQ3\\(trust \&\\robust.)}} \\
    \midrule
    Troiano et al. (Big Data Platform) \cite{bigdataplatformforintegrated} & 2019 & Platform/ architecture (pilots) & \textcolor{Green}{$\checkmark$} & \textcolor{Red}{$\times$} & \textcolor{Red}{$\times$} \\
    Moeckel (Attacker-Centric) \cite{attackercentricthinking} & 2020 & Practitioner interviews & \textcolor{Red}{$\times$} & \textcolor{Red}{$\times$} & \textcolor{Green}{$\checkmark$} \\
    Mahmood \& Abbasi (Phishing forecast) \cite{usingdeepgenerativemodels} & 2020 & Case study (DGMs forecasting) & \textcolor{Green}{$\checkmark$} & \textcolor{Red}{$\times$} & \textcolor{Red}{$\times$} \\
    Cascavilla et al. (CTI SLR) \cite{cybercrimethreatintelligence} & 2021 & Systematic multi-vocal review & \textcolor{Red}{$\times$} & \textcolor{Green}{$\checkmark$} & \textcolor{Red}{$\times$} \\
    M{\"o}ller (Digital transformation) \cite{cyberseucirtyindigitaltransformation} & 2023 & Governance/ processes (monograph) & \textcolor{Green}{$\checkmark$} & \textcolor{Green}{$\checkmark$} & \textcolor{Red}{$\times$} \\
    Darem et al. (CTI taxonomy) \cite{cyberthreatsclassif} & 2023 & Sector review + framework & \textcolor{Red}{$\times$} & \textcolor{Red}{$\times$} & \textcolor{Green}{$\checkmark$} \\
    Humphreys (AI hype \& responsibility) \cite{aihypeasacybersecurityrisk} & 2024 & Governance/ ethics analysis & \textcolor{Red}{$\times$} & \textcolor{Green}{$\checkmark$} & \textcolor{Red}{$\times$} \\
    Deshpande (AI threats in FS) \cite{cybersecurityinfinancialservices} & 2024 & Narrative review (FS) & \textcolor{Red}{$\times$} & \textcolor{Green}{$\checkmark$} & \textcolor{Green}{$\checkmark$} \\
    Devavarapu et al. (Card fraud – outliers) \cite{creditcardfraud} & 2024 & Technical/ experimental & \textcolor{Green}{$\checkmark$} & \textcolor{Red}{$\times$} & \textcolor{Green}{$\checkmark$} \\
    Gong et al. (Network DL) \cite{utilizingdeeplearning} & 2024 & Technical/ experimental & \textcolor{Green}{$\checkmark$} & \textcolor{Red}{$\times$} & \textcolor{Green}{$\checkmark$} \\
    Yu et al. (Fin. network monitoring) \cite{designoffinancialnetwork} & 2024 & Risk/state assess. (Bayesian) & \textcolor{Green}{$\checkmark$} & \textcolor{Red}{$\times$} & \textcolor{Green}{$\checkmark$} \\
    Falowo et al. (Malware \& DDoS trends) \cite{evolvingmalwareandddos} & 2024 & Longitudinal trend study & \textcolor{Red}{$\times$} & \textcolor{Green}{$\checkmark$} & \textcolor{Green}{$\checkmark$} \\
    \bottomrule
  \end{tabularx}
  \caption{Evidence matrix (12 papers) recast by research questions (RQs): RQ1—adoption, RQ2—barriers, RQ3—trust \& robustness. A specific field gets a $\checkmark$ if the theme is explicitly discussed; otherwise $\times$ is given.
}
  \label{tab:ai-criteria-rq}
\end{table}

\subsubsection*{Analysis phase}
The \textbf{12} papers were manually analysed with the goal of answering RQ0 and our adoption-focused questions: whether financial institutions actually use AI within CTI, at which CTI stages (collection, processing, analysis, dissemination) and in what forms, and what barriers are reported. For each study, we assessed evidence of real-world uptake and the presence of domain- or role-based perspectives (e.g., Security Operations Center (SOC) analysts, CTI leads, incident responders, risk/compliance). In addition, we extracted the considered organisations/sectors, sample size, methodology (experiments, case studies, interviews/surveys), and stated objectives/outcomes, and we noted any cues related to analyst trust/explainability, integration and operational fit (e.g., compatibility with SIEM and Security Orchestration, Automation, and Response (SOAR) platforms, latency/cost), and adversarial exposure when available. To this purpose, we created a codebook (which can be found in the repository) through which two authors analysed each paper. We report the inter-code reliability: Cohen’s kappa ($\kappa$)=0.95, which demonstrates strong agreeability~\cite{warrens2015five}.

\subsubsection*{Results}
Let us present the major findings that stem from Table~\ref{tab:ai-criteria-rq} that directly relate to our preliminary RQ.

\begin{itemize}[leftmargin=*,noitemsep,topsep=0pt]
\item \textbf{Capabilities and evidence base.} Among the 12 papers, 33\% report higher accuracy for supervised/deep models than rule-based baselines on finance-relevant CTI tasks (card-fraud screening, phishing-activity forecasting, network/attack detection) \cite{creditcardfraud,usingdeepgenerativemodels,utilizingdeeplearning,designoffinancialnetwork}. A further 17\% contribute CTI lifecycles/taxonomies that provide structure but are not finance-specific \cite{cybercrimethreatintelligence,cyberthreatsclassif}. Finally, 25\% present longitudinal/threat-trend analyses showing \emph{escalating attacker activity}—for instance, rising malware volumes and more frequent/intense DDoS incidents—thereby motivating AI-enabled CTI \cite{evolvingmalwareandddos,cybercrimethreatintelligence,cyberthreatsclassif}.

\item \textbf{Architectures and adoption.} Of the 12 papers, 17~\% propose sector-tailored monitoring/platform designs for financial infrastructures—one explicitly aligned with STIX and its finance profile, FINSTIX\footnote{\textbf{STIX} (Structured Threat Information Expression) is an OASIS standard that defines machine-readable CTI objects (e.g., indicators, TTPs, malware, campaigns) and their relationships for consistent exchange. \textbf{FINSTIX} is a financial-services profile of STIX that adapts objects and vocabularies to banking use (e.g., payment flows, counterparties, market infrastructures) and prescribes sharing workflows among firms and regulators.}—yet only 8~\% report a pilot or operational deployment; overall production evidence remains scarce \cite{bigdataplatformforintegrated,designoffinancialnetwork}.

\item \textbf{Barriers and gaps.} Human/organizational factors are addressed in 17\% (analyst trust, usability, accountability) \cite{attackercentricthinking,aihypeasacybersecurityrisk}; another 17\% analyze AI-specific risks for finance (data poisoning, model misuse/GenAI-enabled deception) \cite{cybersecurityinfinancialservices,aihypeasacybersecurityrisk}. Across the set, the \emph{dominant frictions} include: (i) governance/compliance and model risk management (need for explainability, auditability, and documented controls), (ii) integration with existing security operations tooling and legacy systems (iii) operational constraints (latency, cost/MLOps burden, data access/quality), and (iv) limited evidence on robustness to adversarial manipulation—together highlighted in 25\% under governance/controls \cite{aihypeasacybersecurityrisk,cyberseucirtyindigitaltransformation, cyberthreatsclassif}.
\end{itemize}

\vspace{0.5em}
\noindent
\begin{cooltextbox}
\textbf{Answer to RQ0.} AI for CTI in finance shows promise but little operational proof. \textbf{33\%} of papers report benchmark gains; \textbf{17\%} propose sector-tailored architectures (\textbf{8\%} piloted). Evidence on human factors and on the security \emph{of} AI in CTI is very limited (\textbf{17\%} each). Overall, the production use is rare; trust, integration, and robustness remain open issues.

%Reported \emph{adoption barriers} echo our findings (see §\ref{sec:results}) that governance/compliance constraints, integration \& data-plumbing complexity, limited explainability/trust, and skills represent major challenges.

\end{cooltextbox}
\subsubsection*{Observations}
Taken together, the literature is \emph{benchmark-heavy and operations-light}. Most studies optimise accuracy on public or historical data, with few pilots and scarce reporting on analyst workload, Mean Time To Repair/Recover/Resolve/Respond (MTTR), latency, or integration cost. Human/organizational considerations and finance-specific evaluations of adversarial risk are underexplored. These patterns point to a translation gap from prototypes to practice and motivate our focus on adoption and barriers in real CTI workflows. Finally, we did not identify any other mixed-method studies in our final set of papers.

\subsection{Research Gap and Problem Statement}
\label{ssec:problem}

\subsubsection*{Gap}
Despite some technical progress, it remains unclear \emph{whether} financial institutions are using AI within CTI at scale and \emph{what prevents adoption}. Prior work seldom examines real-world uptake, analyst trust/explainability, regulatory readiness or SOC/SIEM integration. Recent analyses of CTI data ecosystems by Schröer et al. \cite{schroer2025dark} highlight fragmentation, heterogeneous provenance and inconsistent data quality across surface, deep, and dark web sources, which represent factors that can further hinder the trustworthy operational use of AI in financial CTI.

\vspace{-0.3em}
\subsubsection*{Problem statement}
Financial institutions require AI-enabled CTI that is actionable and auditable in production and resilient to adversarial manipulation. Current research demonstrates potential but offers limited guidance for achieving trustworthy, integrated, and robust deployment under financial-sector constraints.
\vspace{-0.3em}
\subsubsection*{Research questions}
\begin{itemize}[leftmargin=*,noitemsep,topsep=0pt]
  \item \textbf{RQ1 (adoption).} \textcolor{violet!90!black} {To what degree are financial institutions using AI for CTI?}
  \item \textbf{RQ2 (barriers).} \textcolor{violet!90!black} {What barriers (technical, regulatory, human, organizational) limit adoption and operational use of AI-enhanced CTI in finance?}
  \item \textbf{RQ3 (trust \& robustness).} \textcolor{violet!90!black} {What factors shape the trust in AI-assisted CTI, and how are adversarial risks (e.g., poisoning, GenAI misuse) managed?}
\end{itemize}

\section{Research Method}
\label{sec:method}
\noindent

\noindent
To investigate how AI affects CTI adoption and operational use in financial cybersecurity, we apply a mixed-methods design combining (i) a systematic literature review (SLR; \S\ref{ssec:slr}), (ii) semi-structured interviews (\S\ref{ssec:interviews}), and (iii) a complementary survey (\S\ref{ssec:surveys}).%
\footnote{SLR followed established guidelines~\cite{kitchenham2015evidence}; interviews followed Kallio et al.~\cite{kallio_systematic_2016}; qualitative analysis used Mayring~\cite{mayring_qualitative_2010}; survey responses were analyzed descriptively.}

\subsection{Semi-structured Interviews}
\label{ssec:interviews}

In the interview stage, we conducted semi-structured interviews with practitioners from cybersecurity departments in finance and finance-focused consulting.
This stage served two purposes: (i) to contextualize and validate signals from the SLR (\S\ref{ssec:slr}) and (ii) to elicit practice-driven mechanisms and constraints that informed the survey design.

\begin{table}[ht]
\footnotesize
\centering
\begin{tabularx}{\linewidth}{|c|X|c|c|c|}
\hline
\textbf{ID} & \textbf{Role / Dept.} & \textbf{Age} & \makecell{\textbf{Exp.} \\ \textbf{(years)}} & \textbf{Domain} \\
\hline
1 & Information Security Officer                      & 36--45 & 8    & Banking \\
2 & Team Leader Cyber Defense Center                  & 36--45 & 4+   & Banking \\
3 & Head of IT Network \& Security Team               & 36--45 & 2    & Banking \\
4 & Information Security Officer                      & 36--45 & 7    & Banking \\
5 & Cybersecurity AI Lead & 26--35 & 4--5 & Consulting \\
6 & Cybersecurity AI Consultant  & 36--45 & 10+  & Consulting \\
\hline
\end{tabularx}
\caption{Interview Participant Characteristics}
\label{tab:interview_data}
\end{table}

\textbf{Description of Companies:}
The interview phase targeted practitioners from four financial institutions (banks) and two consulting firms specializing in financial cybersecurity.
All selected organizations were mid- to large-sized institutions with established security operations. Consulting firms were included to capture cross-institutional perspectives from advisors working with multiple banks (Table~\ref{tab:interview_data}).

\textbf{Interview Design:}
\noindent Participants were recruited based on clear selection criteria, including sector affiliation, seniority, and direct involvement with AI-driven CTI tools. The final interview sample comprised professionals in senior roles such as Head of Cyber Defense, Head of IT Network and Security, Information Security Officer, and finance-focused security consultants (Table~\ref{tab:interview_data}). Each interview lasted 45--60 minutes and followed a semi-structured guide. Thematic blocks included: (i) practical experiences with AI-enhanced CTI deployment, (ii) perceived benefits and limitations of AI tools, (iii) organizational barriers to adoption, and (iv) expectations for future integration. The structure combined guided prompts with open-ended questions to ensure both comparability and flexibility. All interviews followed a guide aligned with the research questions (Appendix~\ref{app:second}).

\textbf{Distribution:}
\noindent Interviews were conducted virtually via Microsoft Teams. Sessions were audio-recorded with consent and transcribed verbatim. All data were collected between February and March 2025. The interviews were exploratory in nature and served as the qualitative foundation for designing the subsequent survey.

\begin{table}[ht]
\footnotesize
\centering
\begin{tabularx}{\linewidth}{|c|X|c|c|c|}
\hline
\textbf{ID} & \textbf{Role} & \textbf{Age} & \makecell{\textbf{Exp.} \\ \textbf{(years)}} & \textbf{Sector} \\
\hline
1  & Chief Information Secruity Officer                         & 46--55 & 6--10  & Finance \\
2  & Information Security Officer  & 36--45 & 6--10  & Finance \\
3  & Financial Officer             & 46--55 & 11--20 & Finance \\
4  & Head of Cyber Defense Center     & 36--45 & 11--20 & Finance \\
5  & Head of IT Security              & 36--45 & 11--20 & Finance \\
6  & Chief Information Secruity Officer                         & 46--55 & 6--10  & Finance \\
7  & Information Security Officer  & 36--45 & 6--10  & Finance \\
8  & Information Security Officer  & 36--45 & 0--2   & Finance \\
9  & Chief Information Security Officer                         & 46--55 & 11--20 & IT \\
10 & SOC Teamlead                 & 36--45 & 6--10  & IT \\
11 & Security Consultant           & 26--35 & 3--5   & IT \\
12 & Head of Security Operations   & 18--25 & 3--5   & Energy \\
13 & Chief Information Security Officer                        & 46--55 & 0--2   & Education \\
14 & Cybersecurity Intern           & 26--35 & 0--2   & Prof. Service \\
\hline
\end{tabularx}
\caption{Participant and Survey Characteristics}
\label{tab:survey_data}
\end{table}

\textbf{Data analysis:}
\noindent A thematic analysis approach was applied using Mayring’s content analysis methodology \cite{mayring_qualitative_2010}. First, transcripts were read in full to identify recurring patterns, challenges, and success factors in AI-driven CTI adoption. These concepts were captured through open coding. In a second step, codes were consolidated into broader categories (corresponding to our RQs), including adoption barriers, trust and explainability, organizational readiness, and regulatory implications. Coding and category refinement were conducted using a predefined extraction schema derived from the interview guide and iteratively updated as new themes emerged. To reduce researcher bias, an independent second researcher reviewed the coding scheme and a subset of the coded transcripts. Inter-coder agreement was assessed and showed high consistency (Cohen $\kappa$=0.93), following established practices in prior transparency and cybersecurity interview studies \cite{braun2024understanding}. This coding process allowed triangulation with findings from the systematic literature review. To ensure traceability, each reported qualitative finding can be linked back to one or more coded interview excerpts. Representative quotes were used to validate and illustrate the identified patterns, while contradictory or minority views were retained to avoid overgeneralization.

\begin{figure*}[t]
    \centering

    \begin{subfigure}[t]{0.32\textwidth}
        \centering
        \includegraphics[width=\textwidth]{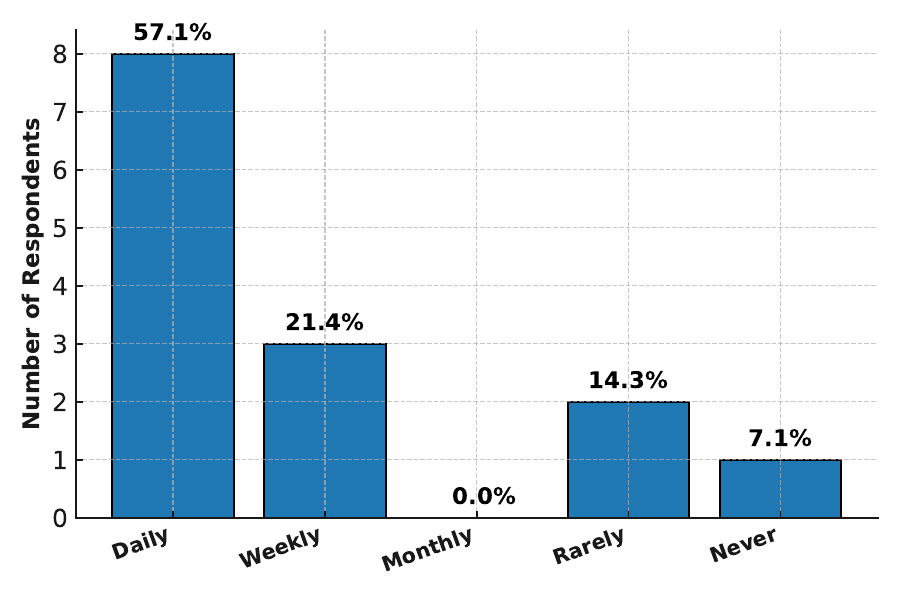}
        \caption{AI use in cybersecurity}
        \label{fig:ai_use}
    \end{subfigure}
    \hfill
    \begin{subfigure}[t]{0.32\textwidth}
        \centering
        \includegraphics[width=\textwidth]{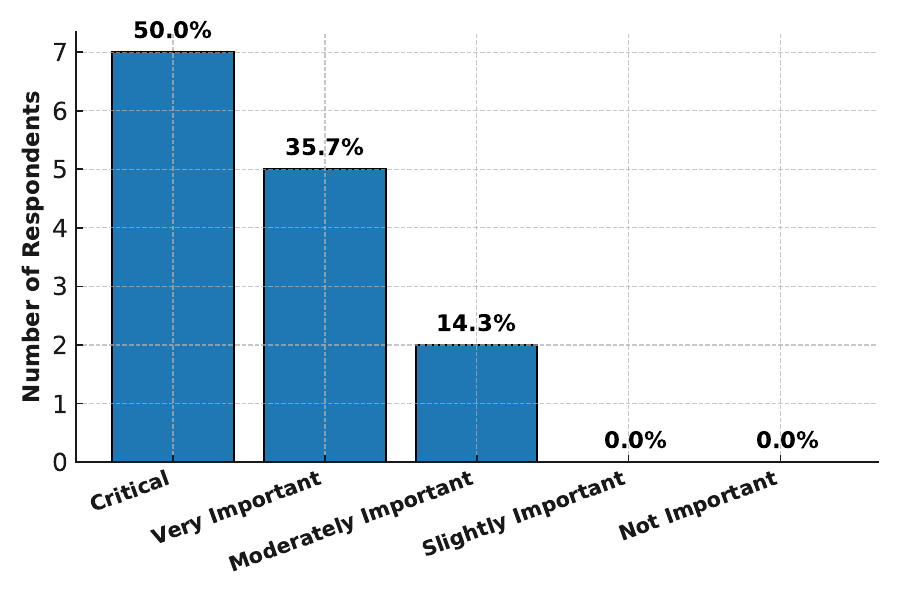}
        \caption{Importance of CTI in cybersecurity}
        \label{fig:cti_importance}
    \end{subfigure}
    \hfill
    \begin{subfigure}[t]{0.32\textwidth}
        \centering
        \includegraphics[width=\textwidth]{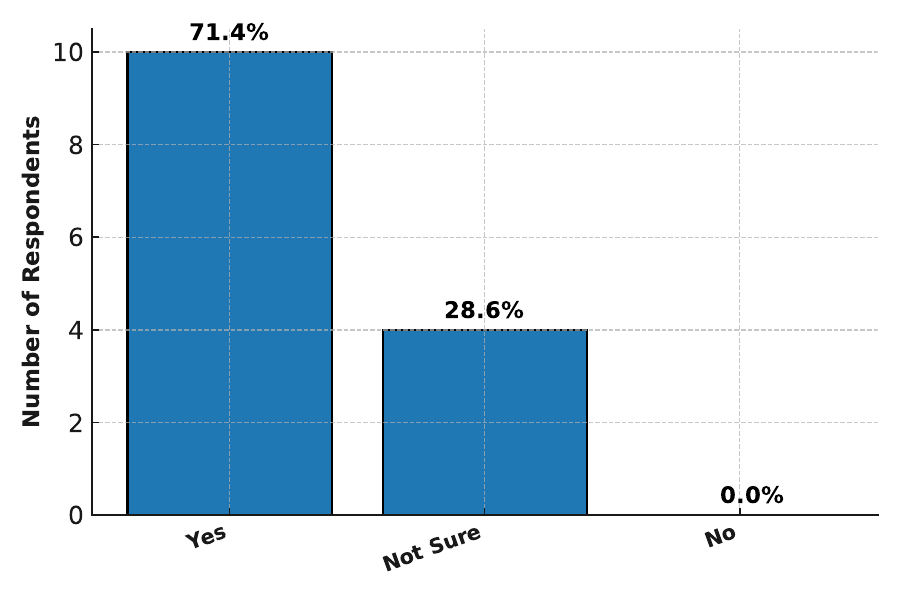}
        \caption{Perceptions of AI dominance in the next five years}
        \label{fig:ai_future}
    \end{subfigure}

    \caption{Adoption of AI in CTI (RQ1).}
    \label{fig:rq1}
    \vspace{-3mm}
\end{figure*}

\textbf{Ethical considerations:}
\noindent Due to the sensitivity of cybersecurity operations, confidentiality was ensured through non-disclosure agreements (NDAs). All interviewees were anonymized, and references to specific institutions or tools were removed from transcripts. Participation was voluntary, and participants were allowed to withdraw at any point. Due to operational restrictions, some responses were deliberately limited, which stresses the ethical and practical challenges of conducting research in this domain.

\subsection{Surveys}
\label{ssec:surveys}

In the survey stage, we conducted a follow-up quantitative survey to validate and extend the findings from the interviews (\S\ref{ssec:interviews}).

\textbf{Description of Companies:}
\noindent To broaden and validate the interview findings, we distributed an online survey across the same financial institutions and consulting firms that participated in the interview phase.\footnote{Compared to interviews, we were able to reach additional personnel within the same organizations.}\footnote{The survey sample also includes a small number of participants from adjacent sectors (Energy and Education). These responses are used only to triangulate governance-style constraints common in regulated/high-impact environments (e.g., auditability, accountability, risk management), and are not used for finance-specific generalization.}
The target group included cybersecurity officers, consultants, and IT security specialists with experience in AI-related tools (Table~\ref{tab:survey_data}).

\textbf{Survey Design:} The survey was structured and distributed using Microsoft Forms (GDPR-compliant) to capture both closed-ended and open-ended responses. Items included Likert-scale questions on the importance of CTI sources, concerns over adversarial AI, trust in AI-driven recommendations, and perceived future challenges such as data bias, explainability, and regulatory compliance. Additional multiple-choice and Likert-based questions assessed respondents’ priorities for awareness training (e.g., deepfake detection, phishing, password hygiene). Open-ended prompt allowed participants to elaborate on their institutional experiences with AI in cybersecurity. More information about the demographics of our survey is provided in the Appendix~\ref{app:second}.

\textbf{Distribution:} The survey was distributed electronically; participation was voluntary and anonymous. We collected 14 complete responses. Given the small, non-random sample and access constraints typical of financial cybersecurity research, we do not claim statistical generalization; instead, results are interpreted as indicative signals that complement the qualitative findings. The voluntary format may have introduced self-selection bias, as respondents more interested in AI were more likely to participate.

\textbf{Data analysis:} Responses were aggregated and analyzed descriptively. Closed-ended items were summarized by frequencies (and relative importance where applicable), while open-ended answers were thematically categorized using Mayring’s approach~\cite{mayring_qualitative_2010}. Most closed-ended items used 5-point Likert scales (e.g., from ``strongly disagree'' to ``strongly agree''). Given the small, non-random sample, we report descriptive statistics only (frequencies and proportions) and intentionally avoid scale aggregation or inferential tests. Quantitative results are interpreted as indicative signals and used primarily to triangulate qualitative mechanisms rather than to estimate prevalence.

\textbf{Ethical considerations:} The survey design adhered to principles of anonymity and confidentiality. No personal identifiers were collected, and participation was strictly voluntary. Respondents were informed about the academic purpose of the study and the exploratory nature of the analysis. We ensured that no institution- or individual-level responses could be traced back to any of the participants.

\section{Results}
\label{sec:results}

\noindent
We first examine the current adoption of AI in CTI within financial institutions (§\ref{ssec:rq1}). 
Then, we analyze the barriers that hinder the deployment of AI-driven CTI (§\ref{ssec:rq2}). 
Lastly, we investigate issues of trust and robustness that shape perceptions of AI in this domain (§\ref{ssec:rq3}). 
All results are based on our mixed-methods approach that combines semi-structured interviews ($n=6$) and a quantitative survey ($n=14$).

\subsection{RQ1: Adoption of AI in CTI}
\label{ssec:rq1}

\noindent \textbf{Interviews:} The findings from interviews showed that the current use of AI in CTI is \textit{limited to specific experiments} rather than being part of everyday practice. An information security officer from a private bank mentioned that they \textit{sometimes test tools that promise machine learning–based anomaly detection, but they have not integrated them into their SOC workflows yet}. This shows how companies are willing to try AI technologies but remain hesitant to embed these tools into established detection and response processes. A team leader from a cyber defense center similarly said that his organization had evaluated vendor solutions with AI features, but \textit{the analysts still rely on traditional correlation and rule-based detection because they cannot depend on outputs that are not transparent}.

\begin{figure*}[t]
    \centering

    \begin{subfigure}[t]{0.32\textwidth}
        \centering
        \includegraphics[width=\textwidth]{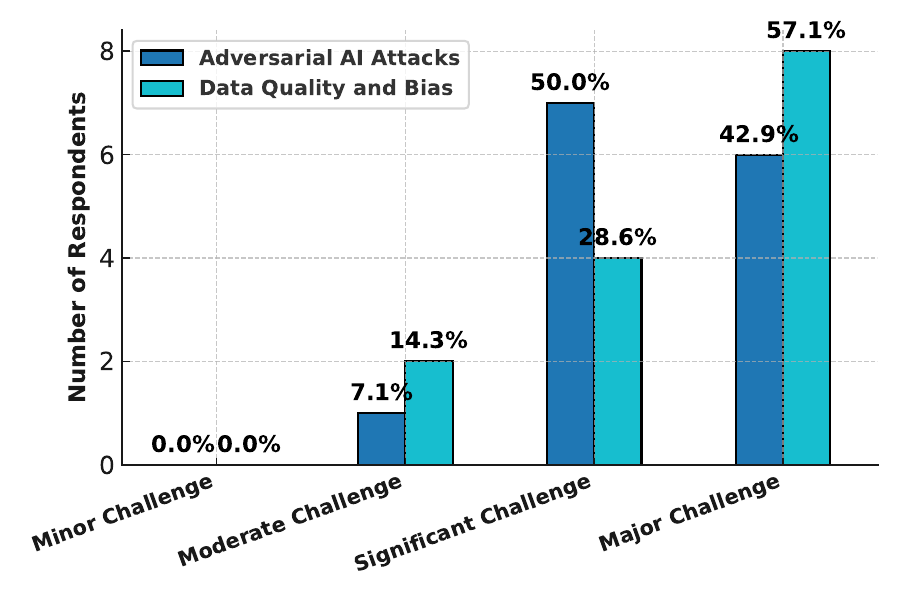}
        \caption{Technical barriers}
        \label{fig:rq2_technical}
    \end{subfigure}
    \hfill
    \begin{subfigure}[t]{0.32\textwidth}
        \centering
        \includegraphics[width=\textwidth]{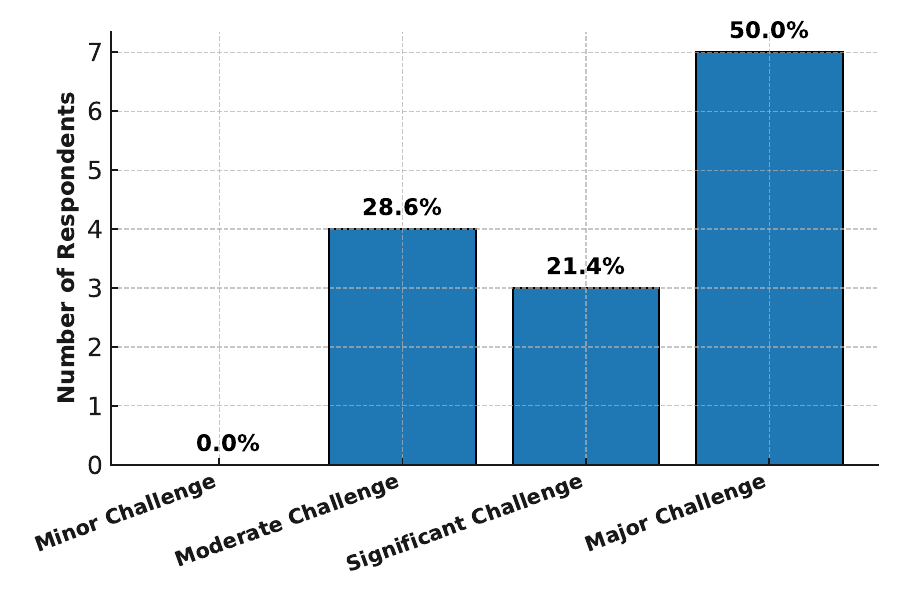}
        \caption{Human barriers}
        \label{fig:rq2_human}
    \end{subfigure}
    \hfill
    \begin{subfigure}[t]{0.32\textwidth}
        \centering
        \includegraphics[width=\textwidth]{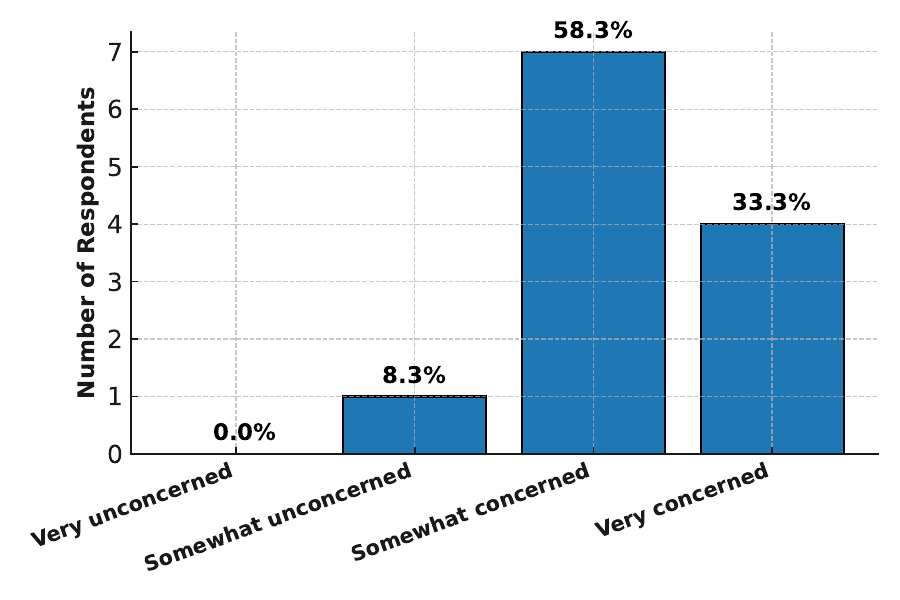}
        \caption{Regulatory barriers}
        \label{fig:rq2_regulatory}
    \end{subfigure}

    \caption{Barriers to adoption of AI-driven CTI (RQ2).}
    \label{fig:rq2_barriers}
    \vspace{-3mm}
\end{figure*}

Financial consultants reported similar experiences when describing the state of AI integration among their clients in the financial sector. One consultant working at a large professional services firm stated that \textit{clients often adopt AI features because they are bundled with existing products, not because they have a clear strategy for how to use them}. He described that institutions often pay for AI-driven modules as part of a SIEM or SOAR package. However, many functionalities remain disabled or are used only in a testing environment. Another consultant stressed that managers are often eager to reference AI in strategic discussions, but \textit{when you look at the day-to-day work of analysts, it is the same dashboards and correlation rules as before}. These accounts reveal a broken link between how AI is positioned in organizational strategy and how it is actually applied and used in operational CTI practice.

The interviews also show that financial institutions mostly see AI as a \textit{tool for exploration} rather than as an operational necessity. It was mentioned that AI-driven services through external providers were provided to these companies. However, they chose not to integrate them, as \textit{they do not want to base decisions on models they do not understand}. Furthermore, it was mentioned that procurement and not the operational demand often determines which tools are available: \textit{if the vendor says the product has AI, then it is there, but that does not mean we will use it}.

\noindent \textbf{Surveys:} The results show that a substantial subset of participants are already making use of AI-driven tools in their current cybersecurity activities. More specifically, 57.1\% reported daily use, 21.4\% indicated weekly use, 14.3\% stated rare use, and 7.1\% reported never using such tools, while no respondents reported monthly use (see Figure~\ref{fig:ai_use}). CTI as a general capability was rated as highly important, with 50.0\% of respondents considering CTI critical, 35.7\% very important, and 14.3\% moderately important, while none rated it as slightly or not important (see Figure~\ref{fig:cti_importance}). \textit{While these numbers indicate that AI applications are present in cybersecurity operations, these are not directly tied to CTI workflows. The high valuation of CTI does not convert to AI-supported CTI practices, which leaves CTI processes unaffected by the reported AI use.} Thus, institutions recognize the significance of intelligence processes in cybersecurity, but AI is not yet uniformly embedded in day-to-day CTI operations.

Expectations regarding the future role of AI are somewhat stronger compared to the current level of implementation. Majority of participants (71.4\%) agreed with the statement that \textit{AI-driven tools will become the dominant approach} in financial cybersecurity within the next five years, while 28.6\% were unsure and none explicitly rejected this prediction (see Figure~\ref{fig:ai_future}). These responses suggest that institutions are acknowledging the \textit{centrality of CTI} and positioning \textit{AI as a future enabler} of that function. However, the lack of demonstrated operational use can lead to a divergence between strategic outlook and current deployment, wherein \textit{CTI is considered highly valuable} but \textit{AI remains peripheral}.

%\noindent \textbf{Surveys:} The results show that only a small subset of participants are making use of AI-driven tools in their current cybersecurity activities. The majority of participants indicated that \textit{adoption is absent} or \textit{limited to exploratory projects}. CTI as a general capability was rated as highly important, with majority of participants stating that it is a central element of their security operations. Yet, the use of AI is still limited within those same operations. Thus, institutions recognize the significance of intelligence processes but \textit{do not yet rely on AI} to enable these.

%Expectations regarding the future role of AI are considerably stronger compared to the current level of implementation. A majority of respondents expressed agreement with the statement that \textit{AI-driven tools will become the dominant approach} in financial cybersecurity within the next five years. Only a minority displayed uncertainty regarding this trajectory, and there were no respondents that explicitly rejected the likelihood of AI gaining a central position. Furthermore, responses further indicated that institutions are acknowledging the \textit{centrality of CTI} and positioning \textit{AI as a future enabler} of that function. However, this is done without showing the real operational use. This leads to a divergence between strategic outlook and current deployment, whrein\textit{ CTI is considered highly valuable} but \textit{AI remains peripheral}.

\begin{cooltextbox}
    \textsc{\textbf{Answer to RQ1:}} Financial institutions currently use AI in CTI only in exploratory ways, mainly through vendor-provided features that are tested but rarely integrated into daily workflows. Analysts continue to rely on rule-based methods due to transparency and trust concerns. Surveys confirm that while participants reported daily and weekly use of AI-driven tools, these applications are not tied directly to CTI workflows. At the same time, CTI was rated as highly important by all respondents. Institutions expect AI to become a key enabler within the next five years, with majority of respondents anticipating a dominant role, but for now it remains peripheral to the CTI lifecycle.
\end{cooltextbox}

%Across these statements, AI emerges less as a trusted capability embedded in workflows and more as a feature that organizations cautiously observe, sometimes purchase, but rarely fully integrate into their cyber defense routines.

\subsection{RQ2: Barriers to Adoption}
\label{ssec:rq2}

\noindent \textbf{Interviews:} Participants that come from banks highlighted issues with \textit{broken infrastructures} and the difficulty of \textit{aligning heterogeneous data sources}, which represents a central barrier for introducing AI into CTI processes. An information security officer stressed that \textit{they have too many data sources that are feeding a model with partial data and resulting in noise}. This means that without access to comprehensive datasets, the outputs of AI models are perceived as unreliable and therefore unsuitable for critical monitoring tasks. Another participant further explained that the underlying issue is not the availability of data but rather its dispersion across legacy systems stating that \textit{you cannot simply put an algorithm on top of ten different log sources and expect something meaningful}. According to their statements, it can be said that technical obstacles are not confined to the models themselves but are very closely linked to the infrastructures in which they are expected to operate.

A second barrier mentioned by interviewees is the lack of expertise at the intersection of cybersecurity and machine learning. Several participants highlighted the lack of experts who are able to interpret AI outputs while also understanding the operational realities of security work. A banking officer stressed that \textit{training existing analysts is difficult, and it is very hard to find staff who really understand both worlds}. Furthermore, it was stated that \textit{hiring people with hybrid profiles is nearly impossible} given the current job market, which makes companies \textit{dependent on external vendors} who may not align their products with internal processes. 

\begin{figure*}[t]
    \centering

    \begin{subfigure}[t]{0.32\textwidth}
        \centering
        \includegraphics[width=\textwidth]{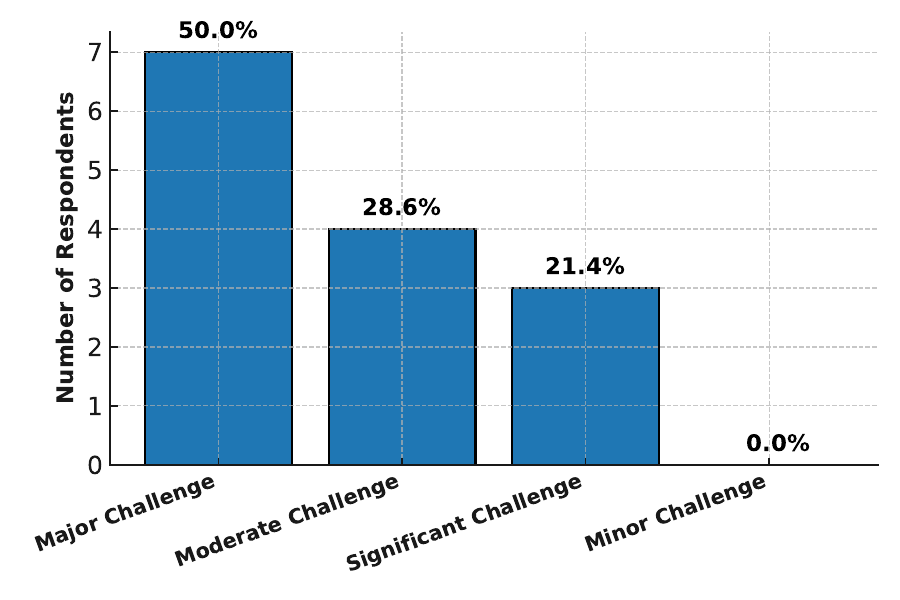}
        \caption{Trust in AI-generated recommendations}
        \label{fig:rq3_trust}
    \end{subfigure}
    \hfill
    \begin{subfigure}[t]{0.32\textwidth}
        \centering
        \includegraphics[width=\textwidth]{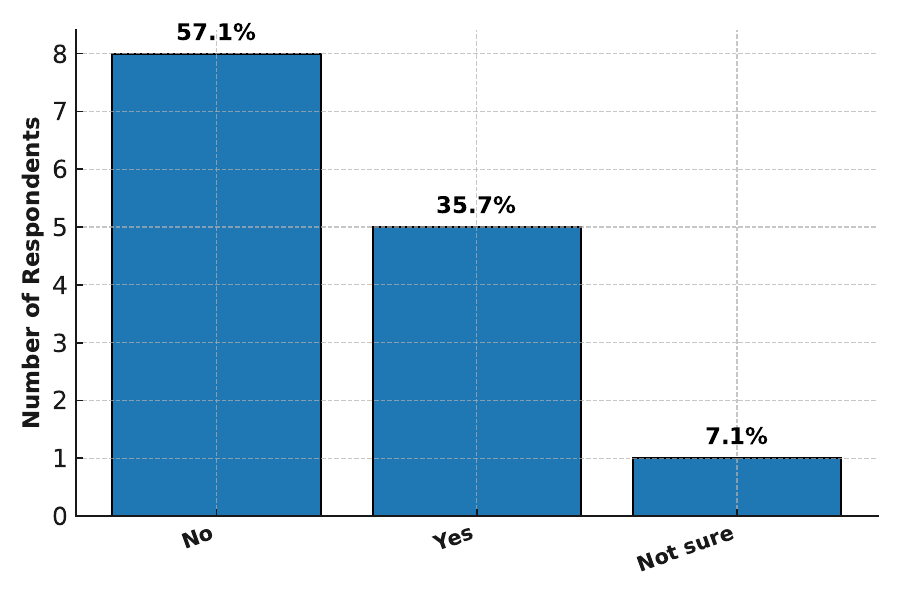}
        \caption{Experience with adversarial AI threats}
        \label{fig:rq3_adversarial}
    \end{subfigure}
    \hfill
    \begin{subfigure}[t]{0.32\textwidth}
        \centering
        \includegraphics[width=\textwidth]{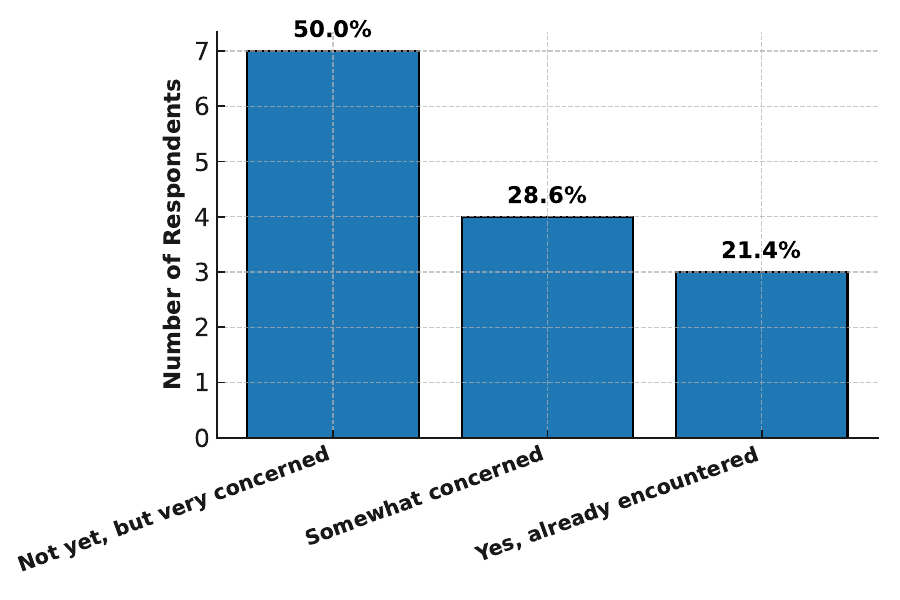}
        \caption{Concerns about misuse of customer-facing AI systems}
        \label{fig:rq3_misuse}
    \end{subfigure}

    \caption{Trust and robustness in AI-driven cybersecurity (RQ3).}
    \label{fig:rq3_results}
    \vspace{-3mm}
\end{figure*}

%These reflections illustrate how workforce limitations hinder the possibility of moving beyond pilot projects, since the necessary interpretative and integration skills are rarely available in-house.

Besides technical and organizational obstacles, participants also noted the influence of \textit{regulatory and audit requirements} in regards to issues in adopting AI-based CTI systems. A security officer mentioned that \textit{if they cannot explain how the system reached a conclusion, their auditors will not accept it}. A team leader similarly underlined that outputs from \textit{AI tools must be justified in internal compliance reviews}, otherwise they cannot be used as the basis for incident response decisions. Thus, regulations are not just outside limits but they actually shape how security teams calculate risks in their daily work. Professionals are cautious about using AI not because they doubt its potential, but because its unclear whether results can be easily matched with the accountability required by auditors inside and outside the organization.

\noindent \textbf{Surveys:} Responses that address challenges in the deployment of AI systems show that \textit{technical obstacles have strong impact}. \textit{Data quality} and \textit{bias} were rated as significant, wherein participants indicated that without reliable and harmonized input data, AI models cannot provide trustworthy results. In fact, 42.9\% of respondents described data-related barriers as a major challenge and 35.7\% as significant (see Figure~\ref{fig:rq2_technical}). In addition, the \textit{fragmentation of data sources} and the \textit{absence of coherent integration} were mentioned as constraints to obtaining value from such tools.

\textit{Human trust in AI recommendations} was also rated as a challenge by most participants, with 50.0\% of respondents assessing \textit{trust in AI} as a major challenge and 28.6\% as significant (see Figure~\ref{fig:rq2_human}) challenge. This is also due to the \textit{lack of analysts} capable of interpreting AI outputs as well as the \textit{skepticism toward automated systems}. Thus, a scarcity of trust among analysts prevents available AI tools from being fully utilized.

Regulatory and governance-related constraints were also mentioned as significant. This includes \textit{compliance with frameworks} such as \textit{GDPR} and the requirement to \textit{justify security decisions in audits}. Participants stressed that \textit{model outputs} must be \textit{defensible in regulatory contexts}. In total, 33.3\% of participants were very concerned and 58.3\% somewhat concerned about regulatory explainability requirements (see Figure~\ref{fig:rq2_regulatory}). They stressed that \textit{model outputs} must be \textit{defensible in regulatory contexts}. Otherwise, their use cannot be justified regardless of achieved technical accuracy.

%\noindent \textbf{Surveys:} Responses that address challenges in the deployment of AI systems show that \textit{technical obstacles have strong impact}. \textit{Data quality} and \textit{bias} were rated as significant, wherein participants indicated that without reliable and harmonized input data, AI models cannot provide trustworthy results. In addition, the fragmantation of data sources and the absence of coherent integration were mentioned as constraints to obtaining value from such tools. 

%\textit{Human trust in AI recommendations} was rated as a challenge by most participants, with many stating that this represents a major concern. This is also due to the \textit{lack of analysts capable of interpreting AI outputs} as well as the \textit{skepticism toward automated systems}. Thus, a lack of trust among analysts prevents available AI tools from being fully utilized.

%Regulatory and governance-related constraints were also mentioned as significant. This includes \textit{compliance with frameworks} such as \textit{GDPR} and the requirement to \textit{justify security decisions in audits}. Participants stressed that \textit{model outputs} must be \textit{defensible in regulatory contexts}. Otherwise, their use cannot be justified regardless of technical accuracy. 

\begin{cooltextbox}
    \textsc{\textbf{Answer to RQ2:}} Three main barriers to adopting AI in CTI were identified: fragmented infrastructures, limited expertise, and regulatory constraints. Companies struggle to integrate heterogeneous data sources, which leads to unreliable outputs. A shortage of professionals who combine cybersecurity and machine learning skills limits the ability to interpret results, which leaves organizations reliant on vendors. Finally, strict audit and compliance requirements mean that AI decisions must be explainable and defensible, which current tools do not fulfill.
\end{cooltextbox}

%The evaluation of these factors underscores the extent to which barriers are located not solely in the models themselves but in the infrastructures and data environments in which they operate.

\subsection{RQ3: Trust and Robustness}
\label{ssec:rq3}

\noindent \textbf{Interviews:} According to interviews, \textit{trust} is considered as the most decisive factor in shaping how practitioners evaluate AI in CTI. Several participants described situations where \textit{models generated alerts that analysts did not understand}, which can lead to a loss of confidence. An information security officer explained that \textit{when the model produces alerts they cannot explain, people stop paying attention to it}. Another participant stated that once \textit{analysts disengage from AI-generated results, the tools quickly become irrelevant in practice}, even if they continue to be part of the technical environment. These insights show that the usefulness of AI is not judged solely by its accuracy but by whether analysts can interpret and act upon its outputs in their operational routines.

Robustness concerns were also highlighted and this is specifically the case when attackers could manipulate AI systems or leverage them offensively. A consultant working on AI risk assessments explained that \textit{we see that attackers already use AI to improve their attacks, while we are still debating how to use it defensively}. The perception that adversaries are one step ahead leads to a cautious attitude among defenders. They often have fear of being locked into systems that might be vulnerable to adversarial manipulation. A team leader at a cyber defense center stated that\textit{ without clear safeguards against data poisoning and model manipulation}, his team would not rely on AI for high-stakes monitoring tasks. Robustness is thus viewed as a precondition for trusting AI in critical infrastructures.

Furthermore, it was highlighted that trust is compromised by the absence of transparent processes that aim to monitor and document AI models over time. A participant explained that when a tool is introduced, \textit{nobody really tells them when it was last retrained or how its performance is measured}. This lack of insight into model evolution and changes creates uncertainty among practitioners, who are aware that models can degrade without continuous oversight. It was also stressed that \textit{without clear documentation, it is impossible to defend the use of AI in audits}, which makes analysts reluctant to engage with such tools. These statements underscore the importance of \textit{building technically sound models} as well as ensuring that they are \textit{accompanied by governance structures} that enable continuous monitoring, retraining, and accountability.

\noindent \textbf{Surveys:} The responses that evaluate human trust in AI recommendations show the \textit{hesitation to fully rely on generated outputs}. Survey data underline this point, as 50.0\% of respondents described trust in AI recommendations as a major challenge and 28.6\% as significant (see Figure~\ref{fig:rq3_trust}). This is the case even among participants that anticipated AI will become dominant in the future. Many participants therefore identified \textit{trust as a barrier} that continues to limit the current use of AI. The persistence of this skepticism indicates that analysts require explainability and reliability as preconditions for any operational adoption.

Responses addressing adversarial threats showed that several participants had already encountered attacks such as \textit{model evasion} or \textit{data poisoning}. In total, 35.7\% reported direct encounters, while 57.1\% were aware of such threats without experiencing them, and 7.1\% indicated no exposure (see Figure~\ref{fig:rq3_adversarial}). Thus, \textit{adversarial manipulation} is widely recognized as a realistic threat, which means that financial institutions cannot view robustness as a theoretical concern and have to treat it as \textit{part of their active defenses}.

In addition, concerns were also raised about the \textit{misuse of AI applications used by customers}. A large share of participants indicated they were concerned about scenarios in which adversaries could \textit{manipulate chatbots or automated systems to disclose sensitive information}. Specifically, 50.0\% stated they were very concerned, 21.4\% somewhat concerned, and 28.6\% reported already having encountered such misuse in practice (see Figure~\ref{fig:rq3_misuse}). These responses show that misuse is not only part of hypothetical discussions but has already manifested in operational environments. The presence of both concerns and reported experiences suggest that trust in customer-facing AI systems is already questioned.

\begin{cooltextbox}
    \textsc{\textbf{Answer to RQ3:}} Interviews and surveys show that trust and robustness are key factors to using AI in CTI. Participants noted that when outputs cannot be explained or applied, tools lose their value, which makes interpretability as important as accuracy. Concerns about adversarial manipulation and data poisoning add contribute to the increased caution, whrein safeguards are seen as essential for high-stakes monitoring. Confidence is also reduced by limited transparency around model retraining and performance tracking, which causes complications with audits.
\end{cooltextbox}

%Thus, concerns over trust and robustness are evident both in relation to adversarial misuse and in the interpretation of outputs within institutional security processes.

%We describe our implementation of the proposed method.

%\textbox{{\small \textbf{Example:} You can also use this textbox by reducing the font-size.}}

%We present the results\footnote{\textbf{Observation.} You can also use footnotes for remarks/observations.}

\section{Discussion}
\label{sec:discussion}
\noindent We first distill the key findings (§\ref{ssec:findings}), then compare them to prior work (§\ref{ssec:comparison}), and finally discuss limitations and implications for practice (§\ref{ssec:limitations}).

\subsection{Key Findings}
\label{ssec:findings}
\noindent Our study reveals that while AI holds significant potential for enhancing CTI in finance, its adoption is hindered by practical realities that extend beyond technical performance. Drawing from the literature review, interviews, and survey, we identify four overarching themes that dominate real-world outcomes: (i) shadow AI in the wild; (ii) a license-first trap; (iii) an attacker perception gap; and (iv) missing security for the AI itself. We synthesize the mixed-method evidence in §\ref{ssec:rq1}--§\ref{ssec:rq3} (including Fig.~\ref{fig:rq1} and the ``Answer to RQ1--RQ3'' summaries) into four recurring deployment patterns.

\noindent \textbf{(i) Shadow AI in the wild.} Practitioners described informal use of public LLMs for fast CTI micro-tasks despite restrictions, creating unvetted data-handling paths and reducing visibility. In finance, this becomes a direct trust barrier because constraints are shaped by confidentiality, auditability, and third-party risk, not only accuracy.

\noindent \textbf{(ii) A license-first trap.} AI adoption is frequently driven by bundled vendor offerings rather than a CTI strategy grounded in concrete workflow needs. As a result, AI features remain disabled or confined to pilots. This aligns with §\ref{ssec:rq1}: institutions often pay for AI modules bundled with SIEM/SOAR products, but functionalities remain disabled or limited to testing, and vendor features are rarely integrated into daily workflows (Answer to RQ1; Fig.~\ref{fig:ai_use}, Fig.~\ref{fig:ai_future}). Procurement-first adoption can also dilute ownership for integration and evaluation, delaying operationalization.

\noindent \textbf{(iii) An attacker perception gap.} Defensive teams sometimes under-weight adversaries' sophisticated use of AI for offensive purposes, such as generating phishing content or adapting to evade detection. This mismatch complicates prioritization and contributes to cautious deployment when teams lack concrete, finance-relevant threat models for AI-enabled CTI pipelines.

\noindent \textbf{(iv) Missing security for the AI itself.} Trust and robustness (RQ3) are undermined by limited drift monitoring, incomplete audit documentation, and insufficient adversarial evaluation. Consistently, trust in AI recommendations was rated as a major challenge by 50.0\% of respondents and as significant by 28.6\% (Fig.~\ref{fig:rq2_human}; Fig.~\ref{fig:rq3_trust}). In addition, 35.7\% reported direct encounters with adversarial AI threats such as model evasion or data poisoning (Fig.~\ref{fig:rq3_adversarial}).

\noindent Collectively, these findings underscore a translation gap. AI's promise in mining data and reducing workload is evident in benchmarks (e.g., 33\% of reviewed papers show accuracy gains), but operational fit in finance requires workflow integration, audit defensibility, and security controls to realize value in a high-stakes environment. This is reflected in the barrier ratings: 42.9\% (major) and 35.7\% (significant) for data-related obstacles (Fig.~\ref{fig:rq2_technical} ), and 33.3\% (very) plus 58.3\% (somewhat) concerned about regulatory explainability requirements (Fig.~\ref{fig:rq2_regulatory}).

\subsection{Comparison to Prior Work}
\label{ssec:comparison}
\noindent Our results extend and contextualize prior research on AI in CTI within a finance domain, which is predominantly prototype-focused as per our SLR. Several studies emphasize benchmark improvements on tasks like fraud detection and phishing forecasting (e.g., \cite{creditcardfraud,usingdeepgenerativemodels,utilizingdeeplearning,designoffinancialnetwork}), but provide limited evidence of sustained production use. Unlike these studies, our user-centric approach highlights why outputs often fail to integrate with analyst workflows or satisfy regulatory audits, echoing governance gaps discussed in \cite{aihypeasacybersecurityrisk,cyberseucirtyindigitaltransformation,cyberthreatsclassif}.

\noindent While \cite{bigdataplatformforintegrated} proposes architectures aligned with STIX, our practitioners report that data integration effort, cost constraints, and unclear ownership frequently prevent operationalization. This aligns with \cite{attackercentricthinking}'s narrative on AI vulnerabilities, while adding practitioner evidence on how these risks are perceived and prioritized. Human factors (only 17\% of papers address them) emerge prominently here: interviews reveal disengagement from unexplained alerts and cautious use when outputs are not audit-defensible. Overall, we bridge the ``bench-to-bedside'' division by providing mixed-methods evidence absent in trend-oriented work that motivates AI use without addressing adoption realities (e.g., \cite{cyberthreatsclassif}).

\subsection{Limitations and Threats to Validity}
\label{ssec:limitations}
\noindent Our study adopts an exploratory design because trustworthy AI deployment in financial CTI is a sensitive, access-restricted domain where large-scale observational or experimental data is rarely available. We discuss key threats and mitigations to contextualize transferability.

\subsubsection*{1) Scope and transferability}
Our evidence comes from a focused, role-diverse cohort (\(n=6\) interviews; \(n=14\) survey responses) across mid-to-large financial organizations (e.g., CISOs, CTI leads, consultants). The sample is in line with practitioner studies in sensitive domains \cite{braun2024understanding}; proportions should be read as indicative signals rather than population estimates. Transferability is strongest for institutions with similar scale, regulatory exposure, and operating models. Self-selection is possible (AI-interested practitioners may have been more willing to participate), which is common in financial-sector interview studies \cite{attackercentricthinking}.

\subsubsection*{2) Temporal coverage and literature scope}
The SLR window (2019--2025) emphasizes peer-reviewed sources and may omit late-2025 developments and gray literature (e.g., vendor reports). We used multi-database searches with reproducible strings to improve coverage\footnote{We used keywords such as ``finance'', ``bank'' and ``financial services''. However, we did not include the keyword ``accounting'' since this can represent many different aspects that can be considered non-finance related, such as ``time accounting''.}, but rapid AI evolution limits any fixed-time snapshot.

\subsubsection*{3) Measurement and response bias}
Survey items are self-reported and concise; constructs such as ``AI use'' and ``adversarial encounter'' capture directional rather than fine-grained frequency or severity. NDAs and operational sensitivities also constrained disclosure depth. We mitigated these via triangulation across the SLR, interviews, and survey; anonymization; and a transparent protocol. Coding/extraction required interpretive analysis and may introduce researcher bias; we addressed this with an independent second review and inter-coder reliability scoring following prior transparency work~\cite{pekaric2025weprovide}. Residual bias may remain.

\noindent \textbf{Implications.} The observed gap between strategic enthusiasm and operational caution suggests value in hybrid deployments (AI with human oversight and clear escalation paths). To reduce risk, standardized robustness checks (e.g., adversarial testing, drift monitoring) and audit-ready documentation should precede scale-up. Future work should quantify operational impact and ROI of pilots, and evaluate socio-technical interventions (workflow integration and governance controls) alongside model performance.

\begin{cooltextbox}
\noindent\textbf{Key takeaways.} First, AI adoption for CTI in finance remains exploratory and often informal, even though most respondents expect rapid growth. Second, the dominant barriers are deployment frictions, especially data integration, skills, and audit defensibility, not model accuracy alone. Third, trust depends on operational transparency and security for the AI itself, including documentation, monitoring for drift, and adversarial evaluation, which practitioners treat as prerequisites before scale-up.
\end{cooltextbox}

\section{Conclusions and Recommendations}
\label{sec:conclusions}

\noindent This study examined the integration of AI capabilities and technologies for CTI in finance through a systematic literature review, practitioner interviews, and surveys. We find that while technical promise is evident---benchmarks show accuracy gains---real-world uptake is rare. Adoption is fragmented and often license-driven rather than needs-driven. Additionally, it is also constrained by trust, regulatory, and integration barriers. Human and organizational factors, along with missing security for the AI itself remain underexplored yet decisive for deployment. In this regard, we conduct (i) a user-centric and mixed-methods evidence base that maps to how AI is integrated in finance across the CTI lifecycle and how it fits workflows and regulatory demands; (ii) we distill four recurring, practice-shaping patterns—shadow AI, license-first adoption, an attacker-perception gap, and absent AI security---with actionable implications for trust and robustness; and (iii) we release practitioner-ready artifacts to enable replication and deployment, including search strings, a coded evidence matrix, and interview/survey instruments.

%\noindent\textbf{Recommendations.} We highlight the following three priorities:  
%1) \textit{Hybrid adoption:} deploy AI to augment, not replace, analysts, pairing automation with explainable oversight.  
%2) \textit{Assurance by design:} mandate drift monitoring, adversarial testing, and audit-ready evidence receipts before activating AI features.  
%3) \textit{Strategic integration:} align adoption with CTI workflows and regulatory defensibility, resisting license-driven enablement in favor of capability-to-control matching.

%In summary, AI can extend CTI’s reach in finance, but its value depends less on raw accuracy and more on trust, governance, and operational fit. Bridging this gap requires joint effort between researchers, practitioners, vendors, and regulators.

\noindent\textbf{Recommendations.}
Grounded in the four practice patterns identified in \S\ref{ssec:findings} and the empirical evidence across RQ1--RQ3, we propose three safeguards that reduce security risk in AI-enabled CTI deployments:
\vspace{-0.5mm}

\begin{enumerate}[leftmargin=*]
    \item \textit{Controlled analyst-in-the-loop operation to prevent misuse and ungoverned ``shadow'' AI.}
    Our findings show that analysts frequently resort to unapproved AI tools under institutional constraints, which creates ungoverned ``shadow'' AI usage and new vectors for data leakage and misuse. In this setting, analyst–AI interaction must be treated as a security boundary rather than a purely usability concern. Concretely, (i) AI outputs used in CTI triage must provide an auditable rationale (e.g., evidence/provenance links and calibrated confidence) to prevent blind reliance on opaque alerts, (ii) low-confidence outputs must default to \emph{human review} rather than automated enforcement actions, and (iii) organizations should provide an approved, monitored pathway for AI assistance (e.g., an internal gateway with redaction and logging) to reduce incentives for unapproved public LLM usage that can leak sensitive incident context.
    \textit{Grounded in:} observed ``shadow AI'' use under institutional restrictions and the recurring trust barrier for adopting AI outputs in operational CTI workflows (see \S\ref{ssec:findings}, RQ1--RQ3).

    \item \textit{Security assurance gates and continuous monitoring for adversarial and operational failures.}
    We observe a gap between practitioner awareness of adversarial and operational risks in AI-enabled CTI and the absence of systematic assurance practices in deployment. As a result, AI components should be treated as attackable systems and subjected to explicit release gates prior to production enablement: evaluation on organization-representative data, robustness checks appropriate to CTI (including adversarial testing where relevant, e.g., evasion/poisoning-style stress tests), and documented acceptance criteria with accountable sign-off. After deployment, continuous monitoring for drift and anomalous behavior, periodic re-evaluation, and rollback procedures need to be implemented. Audit-ready artifacts (model versions, retraining triggers, evaluation summaries, and decision logs) should be maintained so that AI-supported CTI decisions remain defensible under security reviews and regulatory audits.
    \textit{Grounded in:} concerns about adversarial manipulation, limited robustness evaluation in practice, and the lack of audit-ready assurance mechanisms for AI models (see \S\ref{ssec:findings}, RQ3).

    \item \textit{Attack-surface reduction via capability-driven integration and ``disabled-by-default'' vendor features.}
    Practitioner reports indicate that procurement and vendor bundling frequently determine which AI features are enabled. This is done independently of whether those capabilities are operationally integrated into CTI workflows. As a result, a license-driven enablement expands attack surface without corresponding security or operational benefit. To reduce this exposure, it is necessary to enforce capability-driven enablement. Concretely, (i) it should be explicitly specified where AI may influence CTI workflows (e.g., enrichment and prioritization) and where it may not (e.g., autonomous blocking without human approval), (ii) non-integrated AI modules need to be \emph{disabled by default} to avoid unused but reachable functionality, and (iii) it is necessary to map each enabled AI capability to a corresponding control objective (capability-to-control matching), including access restrictions, data-handling constraints, and security logging. This limits unnecessary attack surface and prevents ``license-first'' deployments that increase risk without operational benefit.
    \textit{Grounded in:} license-driven enablement without operational integration and practitioner reports that procurement and bundling shape AI exposure more than capability needs (see \S\ref{ssec:findings}, RQ1--RQ2).
\end{enumerate}

\noindent In summary, trustworthy AI-driven CTI is a security problem as much as a performance problem: deployments must minimize new attack surface, prevent ungoverned usage, and provide continuous assurance and auditability against adversarial manipulation and operational degradation.

% intro 1.5
% Back 2
% method 1.5
% res 2
% discuss 1.5
% concl 0.5
% ref 1

\bibliographystyle{ACM-Reference-Format}

{%\small
\bibliography{bibliography} 
}
% \newpage

\appendices
\section{SLR}
\label{app:first}

\begin{table}[h!]
\centering
\scriptsize
\setlength{\tabcolsep}{5pt}
\renewcommand{\arraystretch}{1.1}
\begin{tabularx}{\linewidth}{@{}>{\raggedright\arraybackslash}X c c c@{}}
\toprule
\textbf{Source} & \textbf{Query Type} & \textbf{Search Results} & \textbf{Included} \\
\midrule
ACM Digital Library & Full-text & 16 & 3 \\
Association for Information Systems (AIS) & Full-text & 50 & 0 \\
IEEE Xplore & Abstract & 116 & 5 \\
NDSS Symposium & Full-text & 2 & 0 \\
Elsevier ScienceDirect & Full-text & 100 & 2 \\
SpringerLink & Full-text & 41 & 2 \\
USENIX Security Symposium & Full-text & 5 & 0 \\
\midrule
\textbf{Total} &  & \textbf{330} & \textbf{12} \\
\bottomrule
\end{tabularx}
\caption{Database Sources, Query Types, and Results for Systematic Literature Review}
\label{tab:slr_search_results}
\vspace{-1em}
{\footnotesize \textit{Note.} Abstract search was used for IEEE Xplore to reduce noise; totals are after de-duplication.}
\end{table}

\begin{table*}[t]
\centering
\scriptsize
\begin{tblr}{
  width = \linewidth,
  rowsep = 0.01pt,
  colsep = 0.1pt,
  colspec = {Q[m,c,0.6cm]Q[m,c,0.6cm]Q[m,l,3.4cm]Q[m,l,8.2cm]Q[m,l,4.4cm]},
  hlines,
  vlines,
  hline{1,2} = {-}{0.08em},
}
\textbf{ID} & \textbf{RQ} & \textbf{Category (Block)} & \textbf{Interview Question} & \textbf{Follow-up / Probes} \\

1 & -- & Background and Role &
What is your current role and position? &
-- \\

2 & -- & Background and Role &
How many years of experience do you have in your role? &
-- \\

3 & -- & Background and Role &
What cybersecurity-related topics are you currently working on? &
-- \\

4 & RQ1 & AI Usage &
How often do you use AI as part of your work? &
Can you give concrete examples? \\

5 & RQ1 & AI Usage &
Can you describe the activities in which you use AI? &
Which tasks benefit most? \\

6 & RQ1 & AI Tools and Adoption &
What specific AI tools do you use in your cybersecurity activities or processes? &
In-house vs. public tools? Adoption factors? \\

7 & RQ1 & AI in Threat Handling &
How do you use AI for threat prevention, identification, detection, or mitigation? &
Specific techniques or workflows? \\

8 & RQ2 & Threat Landscape &
What are the most common cybersecurity threats in the financial sector? &
What makes them unique to finance? \\

9 & RQ2 & AI Benefits and Challenges &
What are the advantages and challenges of using AI to identify, detect, or prevent these threats? &
Comparison to traditional approaches? \\

10 & RQ1 & CTI Usage &
Do you use CTI tools or threat-sharing platforms? &
Are AI capabilities integrated? \\

11 & RQ1 & CTI Data Sources &
Do you use open-source cybersecurity data to train models? &
Which data sources? Is data shared publicly? \\

12 & RQ3 & Future Outlook &
How do you envision the role of AI in financial cybersecurity evolving over the next five years? &
Expected benefits or risks? \\

13 & RQ3 & Adversarial AI &
Have you encountered or do you anticipate adversarial AI threats (e.g., deepfakes, data poisoning, model evasion)? &
Are mitigation strategies in place? \\

14 & -- & Reflection and Closure &
Is there anything else you would like to add that we did not cover? &
Additional concerns or insights? \\

15 & -- & Reflection and Closure &
Do you have any questions or comments regarding this research? &
-- \\

\end{tblr}
\caption{Semi-Structured Interview Guide}
\label{tab:interview_questions}
\end{table*}

\begin{table*}[t]
\centering
\scriptsize
\begin{tblr}{
  width = \linewidth,
  rowsep = 0.01pt,
  colsep = 0.1pt,
  colspec = {Q[m,c,0.6cm]Q[m,c,0.8cm]Q[m,l,3cm]Q[m,l,8.0cm]Q[m,l,4.6cm]},
  hlines,
  vlines,
  hline{1,2} = {-}{0.08em},
}
\textbf{ID} & \textbf{RQ} & \textbf{Category (Block)} & \textbf{Survey Question} & \textbf{Answers} \\

1 & -- & Participant Background &
What is your current role? &
CISO, Information Security Officer, SOC Analyst, Consultant, Intern, Other \\

2 & -- & Participant Background &
Which age group do you belong to? &
18--25, 26--35, 36--45, 46--55, 56+ \\

3 & -- & Participant Background &
What sector do you work in? &
Finance, IT, Consulting, Energy, Education, Other \\

4 & -- & Participant Background &
How many years of experience do you have in cybersecurity? &
0--2, 3--5, 6--10, 11--20, 20+ \\

5 & RQ1 & AI Usage in Cybersecurity &
How often do you use AI-based tools in your cybersecurity work? &
Daily, Weekly, Monthly, Rarely, Never \\

6 & RQ1 & AI Usage in Cybersecurity &
For which tasks does your organization primarily use AI? &
Detection, Analysis, Automation, Reporting, Not used \\

7 & RQ1 & AI Usage in Cybersecurity &
Which AI-based technologies do you use in your organization? &
ML-based detection, LLMs, Vendor AI features, None \\

8 & RQ1 & AI Usage in Cybersecurity &
Does your organization restrict or regulate the use of public AI tools (e.g., ChatGPT)? &
Yes, No \\

9 & RQ1 & Perceived Benefits and Explainability &
AI tools reduce false positives in detection. &
Likert (5-point) \\

10 & RQ1 & Perceived Benefits and Explainability &
AI tools improve the speed of incident response. &
Likert (5-point) \\

11 & RQ3 & Perceived Benefits and Explainability &
AI-generated alerts are sufficiently explainable for operational use. &
Likert (5-point) \\

12 & RQ3 & Perceived Benefits and Explainability &
Lack of explainability limits trust in AI-driven systems. &
Likert (5-point) \\

13 & RQ2 & Threat Landscape and Adversarial Risks &
Which threat types are most relevant to your organization? &
Phishing, Ransomware, Insider threats, AI-enabled threats \\

14 & RQ3 & Threat Landscape and Adversarial Risks &
Has your organization encountered deepfake-related incidents? &
Yes, No \\

15 & RQ1 & CTI Usage and Relevance &
Does your organization use CTI tools or platforms? &
Yes, No \\

16 & RQ1 & CTI Usage and Relevance &
If yes, are these CTI platforms enhanced with AI capabilities? &
Yes, No \\

17 & RQ1 & CTI Usage and Relevance &
What are your primary CTI sources? &
Commercial feeds, ISACs, Open-source, Internal \\

18 & RQ1 & CTI Usage and Relevance &
How important do you consider CTI for your cybersecurity activities? &
Likert (5-point) \\

19 & RQ3 & Threat Landscape and Adversarial Risks &
Have you encountered or are you concerned about misuse of customer-facing AI systems? &
Likert (5-point) \\

20 & RQ1 & Future Outlook &
Will AI-driven tools become the dominant approach in financial cybersecurity in the next five years? &
Yes, No \\

21 & RQ3 & Threat Landscape and Adversarial Risks &
Have you already encountered adversarial AI threats (e.g., evasion, data poisoning)? &
Yes, No \\

22 & RQ2 & Barriers to Adoption &
Adversarial AI attacks represent a major challenge for AI adoption. &
Likert (5-point) \\

23 & RQ2 & Barriers to Adoption &
Data quality and bias represent a major challenge for AI adoption. &
Likert (5-point) \\

24 & RQ2 & Barriers to Adoption &
Regulatory compliance (e.g., GDPR) limits the use of AI in cybersecurity. &
Likert (5-point) \\

25 & RQ2 & Barriers to Adoption &
Human trust in AI recommendations limits adoption. &
Likert (5-point) \\

26 & RQ2 & Future Outlook and Priorities &
Which topics should be prioritized in future cybersecurity awareness training? &
Deepfakes, Phishing, AI misuse, Data protection \\

27 & RQ2--3 & Open Reflection &
Is there anything else you would like to share about your experience with AI in cybersecurity? &
Open-ended \\

\end{tblr}
\caption{Survey Questions and Answers}
\label{tab:survey_instrument}
\end{table*}

\noindent
Table \ref{tab:slr_search_results} reports the per-database results of our SLR, including query type (long/special), total hits, and retained papers after screening.

\section{Interview and Survey Guides}
\label{app:second}

\noindent Table \ref{tab:interview_questions}, we provide the interview guide with all the questions targeting cybersecurity specialists. In addition, in Table \ref{tab:survey_instrument} we provide the questions as well as possible answers from the company survey. The survey was structured into thematic blocks covering participant background, AI usage in cybersecurity, CTI relevance, perceived benefits and limitations of AI, adoption barriers, adversarial risks and future outlook. Survey items were derived directly from themes identified in the semi-structured interviews to support mixed-methods triangulation.

\end{document}